\documentclass[%
 reprint,
nofootinbib,
nobibnotes,
 amsmath,amssymb,
 aps,
]{revtex4-1}
\pdfoutput=1
\usepackage{color}
\usepackage{graphicx}
\usepackage{dcolumn}
\usepackage{bm}
\usepackage{hyperref}
\definecolor{darkred}{rgb}{0.8,0.1,0.1}
\hypersetup{colorlinks=true, linkcolor=darkred, citecolor=blue, linktoc=page}

\begin{document}


\title{Exotic Branes from del Pezzo Surfaces}

\author{Justin Kaidi}
 \email{jkaidi@physics.ucla.edu}
\affiliation{
Mani L. Bhaumik Institute for Theoretical Physics\\
Department of Physics and Astronomy\\
University of California, Los Angeles, CA 90095, USA
}%

\begin{abstract}
We revisit a correspondence between toroidal compactifications of M-theory and del Pezzo surfaces, in which rational curves on the del Pezzo are related to ${1\over 2}$-BPS branes of the corresponding compactification. We argue that curves of higher genus correspond to non-geometric backgrounds of the M-theory compactifications, which are related to exotic branes. In particular, the number of ``special directions" of the exotic brane is equal to the genus of the corresponding curve. We also point out a relation between addition of curves in the del Pezzo and the brane polarization effect.
\end{abstract}

\maketitle


\section{\label{sec:intro}Introduction}

Despite their name, exotic branes are now understood to be ubiquitous and, due to the brane polarization effect, unavoidable even in traditional brane configurations \cite{deBoer:2010ud,deBoer:2012ma}. A distinguishing feature of exotic branes is their unorthodox tension, which can scale as $g_s^{-\alpha}$ with $\alpha>2$ and which has distinguished behavior along some special directions. One denotes by $b_\alpha^{(c_r,\dots,c_2)}$ the exotic brane with tension
\begin{eqnarray}
\nonumber
T_p(b_\alpha^{(c_r,\dots,c_2)}) ={R_{n_1} \dots R_{n_{b-p}} \over g_s^\alpha \,\ell_s^{b+1}}\prod_{k=2}^r\left(R_{m^{(k)}_1} \dots R_{m^{(k)}_{c_k}}\over \ell_s^{c_k} \right)^k
\end{eqnarray}
when the exotic brane worldvolume has $p+1$ non-compact directions (the maximum being $p+1=b+1$), $R_i$ are the radii of the internal directions, and $r \geq 2$.

One way to obtain exotic branes is to wrap traditional, higher-dimensional branes on compactification cycles and apply U-duality transformations to obtain a variety of objects which do not have any clear interpretation in terms of wrapped higher-dimensional branes \cite{Obers:1998fb}. An interesting question is to determine what exactly these exotic branes \textit{do} correspond to in the uncompactified theory.

It has been argued \cite{deBoer:2010ud,deBoer:2012ma} that exotic branes have a higher-dimensional origin as non-geometric backgrounds such as T-folds and U-folds, which extend the traditional notion of a manifold to allow for duality transformations in the transition functions between patches \cite{Hull:2004in,Hull:2006va,Kumar:1996zx,Liu:1997mb,Lu:1998sx}. The fact that exotic branes involve non-trivial T- or U-duality monodromy means that they cannot be realized by any globally well-defined solutions of supergravity. Indeed, there may be configurations for which there is not even a \textit{local} description in supergravity \cite{deBoer:2012ma,Otsuki:2019owg}. Instead, a complete description of these objects is expected to arise only after a manifestly T- or U-duality invariant formulation of the underlying theory has been obtained. One such formulation which has led to considerable progress is Double Field Theory \cite{Hull:2009mi,Aldazabal:2013sca} and its generalization to Exceptional Field Theory \cite{Hohm:2013pua,Hohm:2013vpa,Hohm:2013uia}. Exotic branes have been successfully studied in these formalisms in e.g. \cite{Blair:2014zba,Bakhmatov:2017les,Berman:2018okd,Fernandez-Melgarejo:2018yxq,Kimura:2018hph,Otsuki:2019owg}.

There is however another tentative U-duality invariant formulation of toroidal compactifications of M-theory, known as the ``mysterious duality" \cite{Iqbal:2001ye,HenryLabordere:2002dk,HenryLabordere:2003rd,HenryLabordere:2002xh}. In this ``duality," M-theory compactified on a rectangular torus $T^k$ with vanishing $C$-field vev is put into correspondence with  the del Pezzo surface dP$_k$. In addition to a map between moduli spaces, the correspondence suggests the identification of the spectrum of rational curves on dP$_k$ (subject to some constraints to be reviewed below) with the spectrum of ${1 \over 2}$-BPS branes in the $T^k$ compactification. One might naturally wonder whether the higher genus curves in dP$_k$ also correspond to something in string theory. We will argue below that they correspond to non-geometric backgrounds, which can give rise to exotic branes upon further compactification.

\section{The Mysterious Duality}
We begin by briefly reviewing the correspondence between $T^k$ compactifications of M-theory and the $k$-th del Pezzo surface dP$_k$. First, some basic facts about these surfaces will be needed. Recall that dP$_k$ is $\mathbb{P}^2$ blown up at $k$-points, $k \leq 8$. For $\mathbb{P}^2$ there is a unique class of a curve $H \in H_2 (\mathbb{P}^2, \mathbb{Z})$. Since two lines generically intersect at a point, one has the intersection pairing $H \cdot H = 1$. With each blow-up, one adds an exceptional curve $E_i$ such that $H_2 (dP_{k}, \mathbb{Z}) = \{ H, E_1, \dots, E_k\}$, with the intersection pairings
\begin{eqnarray}
H\cdot H = 1 \hspace{0.3 in} H\cdot E_i = 0 \hspace{0.3 in} E_i \cdot E_j = -\delta_{ij}
\end{eqnarray}
The canonical class for these surfaces is given by $K_{dP_k} = -3H + \sum_{i=1}^k E_i$.

On the other hand, Type IIB is put into correspondence with the Hirzebruch surface $\mathbb{F}^0 = \mathbb{P}^1 \times \mathbb{P}^1$, with compactification on $T^k$ again corresponding to the $k$ point blow-up, which we denote $\mathbb{F}^k$. The 2-cycles corresponding to the two $\mathbb{P}^1$ in $\mathbb{F}^0$ are denoted by $\ell_1$, $\ell_2$. We may think of one of the $\mathbb{P}^1$ as being trivially fibered over the other - since two fibers should not intersect one other, but each intersects the base once, we conclude that $\ell_i \cdot \ell_j  = 1 - \delta_{ij}$. Each blow-up adds an exceptional curve $e_a$, with the intersection pairings
\begin{eqnarray}
\ell_i \cdot \ell_j  = 1 - \delta_{ij} \hspace{0.3 in} \ell_i \cdot e_a = 0 \hspace{0.3 in}e_a \cdot e_b = - \delta_{ab}
\end{eqnarray}
The canonical class for these surfaces is $K_{\mathbb{F}^k} = -2\ell_1 - 2\ell_2 + \sum_{a=1}^{k} e_a$.

It is well known that $\mathbb{F}^k \cong$ dP$_{k+1}$; this is the del Pezzo analog of T-duality \cite{Iqbal:2001ye}. In particular, the bases of 2-cycles for the two surfaces are related by
\begin{eqnarray}
\label{Tduality}
H \mapsto \ell_1 + \ell_2 - e_1  \hspace{0.2 in} E_1 \mapsto \ell_2 - e_1 \hspace{0.2 in} E_2  \mapsto \ell_1 - e_1\,\,\,
\end{eqnarray}
as well as $E_{a+1} \mapsto e_a$ for the remaining exceptional curves.
Furthermore, the invariance of $\mathbb{F}^0$ under exchange of $\ell_1$ and $\ell_2$ is interpreted as the invariance of Type IIB under S-duality.

In what sense do the del Pezzo surfaces ``correspond" to the M-theory compactifications? First, there exists a map between the moduli spaces on the two sides. For rectangular M-theory compactifications with zero $C$-field vev, the moduli are the $k$ radii $R_i$ of the rectangular torus $T^k$ and the 11-dimensional Planck length $\ell_p$. These should be identified under those elements of the U-duality group which preserve the rectangular torus and vanishing $C$-field conditions, i.e. the Weyl group $\mathcal{W}(E_k) \subset E_{k(k)}(\mathbb{Z})$. The moduli space is then $\mathbb{R}_+^{k+1} /\mathcal{W}(E_k)$. For the del Pezzos,  the moduli are the $k+1$ ``volumes" $\left\{\omega(H), \omega(E_i) \right \}$ of the 2-cycles,\footnote{The generalized K{\"a}hler class $\omega \in H^2(dP_k, \mathbb{R})$ is not unique and can in fact be shifted such that it is orthogonal to the canonical class. Indeed, we may decompose $\omega = \omega_\perp + \lambda K$ such that $\omega_\perp \cdot K  = 0$. The choice of parameter $\lambda$ corresponds to a choice of scale \cite{Iqbal:2001ye}.} subject to identification by those diffeomorphisms which leave the canonical class invariant. It is a remarkable fact that such diffeomorphisms are given by $\mathcal{W}(E_k)$ for dP$_k$, thus reproducing the moduli space found before. The dictionary between the moduli proposed in \cite{Iqbal:2001ye} is \footnote{For simplicity, we neglect factors of $2 \pi$ in formulas for volumes/tensions. Also, note that for $\mathbb{F}^0\leftrightarrow\,$Type IIB the map is 
\begin{eqnarray}
\label{IIBrule}
\omega(\ell_1) \leftrightarrow - 2 \log \ell_s \hspace{0.4 in} \omega(\ell_2) \leftrightarrow - 2 \log \ell_s- \log g_s
\end{eqnarray}
}
\begin{eqnarray}
\label{rule1}
\omega(H) \leftrightarrow - 3 \log \ell_p \hspace{0.5 in} \omega(E_i) \leftrightarrow - \log R_i
\end{eqnarray}

Even more remarkable is the fact that the ${1 \over 2}$-BPS branes in M-theory compactifications are in one-to-one correspondence with rational (i.e. genus zero) curves in the del Pezzo. Given such a rational curve $\mathcal{C} \in H_2 (dP_k, \mathbb{Z})$, we may calculate the tension $T_p$ and the worldvolume dimension $p+1$ of the corresponding brane via 
\begin{eqnarray}
\label{rule2}
T_p = \mathrm{exp}\, \omega(\mathcal{C})  \hspace{0.5 in} p+1 = d_\mathcal{C}
\end{eqnarray}
In the above, $d_\mathcal{C}$ is the degree of the curve $\mathcal{C}$, defined as its intersection with the anticanonical class, 
\begin{eqnarray}
d_\mathcal{C} = - K_{dP_k} \cdot \mathcal{C}
\end{eqnarray}
For future reference, we note that the \textit{adjunction formula}  \cite{griffithsp} allows us to relate $d_\mathcal{C}$ to the self-intersection and genus $g$ of $\mathcal{C}$ as follows, 
\begin{eqnarray}
\label{adjunction}
\mathcal{C} \cdot \mathcal{C} = 2\, g(\mathcal{C}) - 2 + d_\mathcal{C}
\end{eqnarray}

As an example of this ``duality," consider uncompactified M-theory, which corresponds to $\mathbb{P}^2$. The latter has two genus zero curves $H$ and $2H$, and a canonical class $K = - 3 H$. Using the rules outlined in (\ref{rule1}) and (\ref{rule2}) we conclude that $H$ corresponds to a 2-brane of tension ${ \ell_p^{-3}}$, while $2H$ corresponds to a 5-brane of tension  ${\ell_p^{-6}}$. These correspond to the familiar M2 and M5 branes. We also see an example of a general phenomenon - branes which are electromagnetically dual correspond to rational curves $\mathcal{C}_{e}$ and $\mathcal{C}_{m}$ which are Serre dual, i.e.
\begin{eqnarray}
\label{EMdual}
\mathcal{C}_{e} + \mathcal{C}_{m} = - K
\end{eqnarray}

This same procedure was repeated for Type IIA/B and compactifications thereof in \cite{Iqbal:2001ye}, with agreement in all cases. For the uncompactified case, we list the set of rational curves and their corresponding tensions and string theory interpretations in Table I. Note importantly that we have required $0\leq d_\mathcal{C}\leq 10$ to get a physical interpretation for the corresponding objects.

\begin{table}[htp]
\label{table1}
\begin{minipage}{.54\linewidth}
\begin{center}
\begin{tabular}{|c|c|c|}
\hline
Curve class & $T$ & Type IIA
\\\hline\hline
$E$ & $\ell_s^{-1} g_s^{-1}$ & D0
\\\hline
$H-E$ & $ \ell_s^{-2}$ & F1
\\\hline
$H$ & $ \ell_s^{-3}g_s^{-1}$ & D2
\\\hline
$2H-E$ & $ \ell_s^{-5}g_s^{-1}$ & D4
\\\hline
$2H$ & $ \ell_s^{-6}g_s^{-2}$ & NS5A
\\\hline
$3H-2E$ & $ \ell_s^{-7}g_s^{-1}$ & D6
\\\hline
$4H-3E $& $ \ell_s^{-9} g_s^{-1}$ & D8
\\\hline
\end{tabular}
\end{center}
\end{minipage}%
\begin{minipage}{.45\linewidth}
\begin{center}
\begin{tabular}{|c|c|c|}
\hline
Curve class & $T$ & Type IIB
\\\hline\hline
$\ell_1$ & $\ell_s^{-2} $ & F1
\\\hline
$\ell_2$ & $ \ell_s^{-2} g_s^{-1}$ & D1
\\\hline
$\ell_1+\ell_2$ & $ \ell_s^{-4}g_s^{-1}$ & D3
\\\hline
$2\ell_1 + \ell_2$ & $ \ell_s^{-6}g_s^{-1}$ & D5
\\\hline
$\ell_1+ 2 \ell_2$ & $ \ell_s^{-6}g_s^{-2}$ & NS5B
\\\hline
$3 \ell_1 + \ell_2$ & $ \ell_s^{-8}g_s^{-1}$ & D7
\\\hline
$\ell_1 + 3 \ell_2$ & $ \ell_s^{-8} g_s^{-3}$ & $7^0_3$
\\\hline
\end{tabular}
\end{center}
\end{minipage}%
\caption{Rational curves for dP$_1$ and $\mathbb{F}^0$, together with their string theory interpretations.}
\end{table}%

\section{\label{sec:Exotic Branes}Exotic Branes}

In Table I we see already the appearance of an exotic brane, albeit a rather familiar one - the exotic brane $7^0_3$, also known as the NS7 brane. Exotic branes will appear as rational curves whenever they are codimension-2 in the compactification, as is the case for $7^0_3$ in uncompactified Type IIB.  Instead of trying to catalogue all exotic branes obtained via compactification, our goal will be to obtain all 10- and 11-dimensional non-geometric backgrounds which reduce to the exotic branes upon appropriate compactification. We will see that these non-geometric backgrounds correspond to higher genus curves in the del Pezzos. Our strategy will be to compactify the theory, identify some codimension-2 exotic branes, translate these branes to curves in the del Pezzo, and then take the blow-down (i.e. decompactification) limit. 

Note that if $r=r'$ is the largest number for which the index $c_{r'}$ in $b_\alpha^{(c_r,\dots,c_2)}$ is non-zero, we will refer to the exotic brane as an $(r'-1)$-exotic brane. All traditional branes are 0-exotic. Unless otherwise specified, we will be working with only 0- and 1-exotic branes here. A more thorough analysis will be presented in upcoming work \cite{toappear}.

\subsection{M-theory and Type IIA}
Let us begin by compactifying M-theory or Type IIA on $T^4$ or $T^3$ to $d = 7$. From the string theory point of view, it is known that there are 20 codimension-2 $1\over 2$-BPS objects in $d = 7$, including thirteen 1-exotic branes and seven wrapped D-branes \cite{deBoer:2012ma}. In the del Pezzo picture, these must correspond to rational curves $\mathcal{C}\in H_2(dP_4, \mathbb{Z})$ such that $d_{\mathcal{C}}=5$. A generic element of $H_2(dP_4, \mathbb{Z})$ may be written as $\mathcal{C} = m H -n E- \sum_{i=1}^3 n_i E_i$. Note that we have singled out one of the exceptional curves $E$ -  from the perspective of M-theory $E$ is on equal footing with the other $E_i$, but from the Type IIA perspective $E$ will be taken to be the M-theory circle, i.e. the second element of $H_2(dP_1, \mathbb{Z})$. Making use of (\ref{adjunction}), we have the following constraints on $\{m,n,n_i \}$, 
\begin{eqnarray}
3m - n - \sum_{i=1}^3 n_i  = 5 \hspace{0.2 in} m^2 - n^2 -  \sum_{i=1}^3 n_i^2 = 3\,\,\,\,
\end{eqnarray}
One finds 20 integers solutions to these constraints. The resulting curves and the corresponding codimension-2 objects in $d=7$ M-theory and Type IIA are shown in Table II. We see that we have obtained a number of exotic objects, including e.g. the Kaluza-Klein monopole KK5A, which in standard notation is $5_2^1$A.

\begin{table*}[htp]
\begin{center}
\begin{tabular}{|c|c|c|c|}
\hline
Curve class & $T_4$ & M-theory & Type IIA
\\\hline\hline
$2H-E_i$ & $R_i \ell_p^{-6} \sim R_i \ell_s^{-6} g_s^{-2}$ & M5 $(\mathbf{3})$ & NS5 $(\mathbf{3})$
\\\hline
$2H-E$ & $R_M \ell_p^{-6} \sim  \ell_s^{-5} g_s^{-1}$ & M5$ (\mathbf{1}) $& D4 $(\mathbf{1})$
\\\hline
$3H-E_i - E_j - 2 E_k $& $R_i R_j R_k^2 \ell_p^{-9} \sim R_i R_j R_k^2 \ell_s^{-9}g_s^{-3}$ & KK6 $(\mathbf{3})$ & $6_3^1 (\mathbf{3})$
\\\hline
$3H-2 E - E_i -E_j $& $R_i R_j R_M^2 \ell_p^{-9} \sim R_i R_j \ell_s^{-7}g_s^{-1}$ &KK6 $(\mathbf{3})$ & D6 $(\mathbf{3})$
\\\hline
$3H - E- E_i - 2 E_j $& $R_M R_i R_j^2 \ell_p^{-9} \sim R_i R_j^2 \ell_s^{-8}g_s^{-2}$ &KK6 $(\mathbf{6}) $& KK5A$(\mathbf{6})$
\\\hline
$4H- E - 2 E_1 - 2E_2 - 2E_3$& $ R_M (R_1 R_2R_3)^2 \ell_p^{-12} \sim (R_1 R_2R_3)^2 \ell_s^{-11}g_s^{-3}$ & $5^3 (\mathbf{1})$ & $4_3^3 (\mathbf{1})$
\\\hline
$4H-2E - E_i - 2 E_j - 2 E_k$ & $  R_i (R_M R_j R_k)^2 \ell_p^{-12} \sim R_i (R_j R_k)^2 \ell_s^{-10} g_s^{-2}$ & $ 5^3 (\mathbf{3})$ & $5_2^2$A $(\mathbf{3})$
\\\hline
\end{tabular}
\end{center}
\caption{Rational curves for dP$_4$ and their $d=7$ M-theory and Type IIA interpretations. The multiplicities of configurations are shown in bold. Note that all of the branes above are codimension-2. So for example when we write D6, we really mean a twice-wrapped D6. The choices of wrapped coordinates give the expected multiplicity.}
\label{table2}
\end{table*}%

Having identified these exotic branes with particular curves in the del Pezzo, we would now like to understand where they originate from in the uncompactified theory. From the del Pezzo point of view, the decompactification limit simply corresponds to blowing down the exceptional curves. Beginning with the case of Type IIA, we blow down the $E_i$ and see that $5_2^2$A descends from the curve $4H-2 E \in H_2(dP_1, \mathbb{Z})$, the $4_3^3$ descends from $4H-E$, the KK5A descends from the anticanonical class $-K_{dP_1} = 3H-E$, and the $6_3^1$ descends from $3H$. These curves were not included in Table I since they are not rational - rather they have genus 2, 3, 1, and 1 respectively.  It is these higher genus curves that we claim correspond to the non-geometric backgrounds underlying exotic branes.

Because we began by considering only rational curves in $d=7$, we have obtained only a subset of the possible non-geometric backgrounds. We may continue this analysis by compactifying to lower dimensions $d \geq 3$, leading to the following general conclusion. For a curve $\mathcal{C}_{mn} = m H - n E$ on dP$_1$, we first define the exotic brane non-compact worldvolume dimension as 
\begin{eqnarray}
b_{\mathcal{C}} +1 = d_{\mathcal{C}} - 2 g (\mathcal{C}) 
\end{eqnarray}
which is the generalization of (\ref{rule2}) to higher genus. Then for every curve $\mathcal{C}_{mn}$ (subject to restrictions given momentarily) we obtain a non-geometric background, 
\begin{eqnarray}
\label{IIAResult}
\mathcal{C}_{mn}\in H_2(dP_1, \mathbb{Z})  \,\, \leftrightarrow \,\, \left[ \left(b_{\mathcal{C}_{mn}}\right)_{m-n}^{g(\mathcal{C}_{mn})} \right]\,\,\,\,\,
\end{eqnarray}
The brackets on the right-hand side indicate that we are really referring to equivalence classes of higher-dimensional non-geometric backgrounds, labeled by their lower-dimensional exotic branes. The elements in each equivalence class have the same del Pezzo origin - for example, one has $[7_3^{(1,0)}] \sim [4_3^3]$ in Type IIA and $[7^{(2,0)}_3] \sim[4_3^{(1,3)}] \sim [1_3^6]$, $ [6_3^{(1,1)}] \sim [3_3^4]$ in Type IIB.

With the map (\ref{IIAResult}), we may now classify all backgrounds - geometric and non-geometric - which descend to ${1 \over 2}$-BPS traditional and 1-exotic branes in Type IIA. These are labeled by $\mathcal{C}_{mn}$ with $m,n \in \mathbb{Z}$ such that 
\begin{eqnarray}
\label{restriction}
&\vphantom{.}&b_{\mathcal{C}_{mn}} +1 \geq 0\hspace{0.3 in} g (\mathcal{C}_{mn}) \geq 0
\nonumber\\
&\vphantom{.}&b_{\mathcal{C}_{mn}} +g (\mathcal{C}_{mn}) + 1 \leq 10
\end{eqnarray}
 We list all possibilities in Table III, indexed by the power  $\alpha$ of $g_s$ in the tension formula, $T \sim g_s^{-\alpha}$. In the current case, we have $\alpha = m-n$. 

\begin{table}[htp]
\begin{center}
\begin{tabular}{|c|c|c|}
\hline
$\alpha$ & Type IIA &  Type IIB
\\\hline\hline
0 & $[1_0^0]$ & $[1_0^0]$
\\\hline
1 & $[0_1^0],\, [2_1^0],\, [4_1^0],\, [6_1^0],\, [8_1^0]$ & $[(-1)_1^0], [1_1^0], \,[3_1^0],\, [5_1^0], \,[7_1^0],\, [9_1^0]$
\\\hline
2 & $[5_2^\ell$A]  \,\,$\ell\in \{0,\dots,4\}$& $[5_2^\ell$B] \,\, $\ell\in \{0,\dots,4\}$
\\\hline
3 & $[6_3^1], \,[4_3^3],\, [2_3^5], \,[0_3^7] $& $[7_3^0],\, [5_3^2], \,[3_3^4],\, [1_3^6],\, [(-1)_3^8]$
\\\hline
4 & $[9_4^0$A], [$5_4^3$A], [$1_4^6$A]& [$9_4^0$B], [$5_4^3$B], [$1_4^6$B]
\\\hline
5 & $[2_5^6]$& $[5_5^4],\, [(-1)_5^8]$
\\\hline
6 & $-$& $-$
\\\hline
7 & $[0_7^9]$& $-$
\\\hline
\end{tabular}
\end{center}
\caption{(Non-)geometric backgrounds descending to (exotic) branes in Type IIA/B with tension scaling as $T \sim g_s^{-\alpha}$. }
\label{table3}
\end{table}%

Before moving on to Type IIB, we first describe the situation for M-theory. Decompactification now involves blow-down of $E$ as well. The exotic brane $5^3$ is then seen to descend from the curve $4H \in H_2(\mathbb{P}^2, \mathbb{Z})$, while the KK6 descends from the anticanonical class $-K_{\mathbb{P}^2} =3H$. These curves are again not rational - they are of genus $3$ and $1$, respectively. More generally, given a curve $\mathcal{C}_n = n H$, we claim that there is a corresponding class of non-geometric backgrounds in M-theory,
\begin{eqnarray}
\mathcal{C}_n \in H_2(\mathbb{P}^2, \mathbb{Z}) \,\, \leftrightarrow \,\,\left[\left( b_{\mathcal{C}_{n}} \right)^{g(\mathcal{C}_n)}\right]
\end{eqnarray}
 labelled by 1-exotic branes so long as (\ref{restriction}) is satisfied.\footnote{Note that the right-hand side of the final line of (\ref{restriction}) is now $11$.} In fact, these conditions restrict us to $n \in \{1, \dots, 5 \}$, which give five corresponding (non-)geometric backgrounds $[2^0]$, $[5^0]$, $[6^1]$, $[5^3]$, and $[2^6]$. This reproduces the standard M2 and M5 branes, the KK6, and the two known 1-exotic brane backgrounds \cite{Bakhmatov:2017les,Fernandez-Melgarejo:2018yxq}.

\subsection{Type IIB}

We may repeat the above exercise for Type IIB. We begin by considering the theory compactified on $T^3$. The spectrum of ${1\over 2}$-BPS codimension-2 objects may be obtained by T-dualizing the Type IIA spectrum presented above. Recall that at the level of the del Pezzo, T-duality just amounts to a change of basis 2-cycles, as shown in (\ref{Tduality}). Implementing this change of basis and using (\ref{IIBrule}), we obtain the spectrum of curves and corresponding branes shown in Table IV.

\begin{table}[htp]
\begin{center}
\begin{tabular}{|c|c|c|}
\hline
Curve class & $T_4$ &  Type IIB
\\\hline\hline
$\ell_1 + 2 \ell_2 - e_i$ & $ R_i \ell_s^{-6} g_s^{-2}$ & NS5 $(\mathbf{3})$
\\\hline
$2\ell_1 + \ell_2 - e_i$ & $ R_i \ell_s^{-6} g_s^{-1}$ & D5 $(\mathbf{3})$
\\\hline
$2 \ell_1 + 2 \ell_2 - 2 e_i - e_j$& $R_i^2 R_j \ell_s^{-8}g_s^{-2}$& KK5B $(\mathbf{6})$
\\\hline
$3 \ell_1 + \ell_2 - e_1 - e_2 - e_3$& $R_1 R_2 R_3 \ell_s^{-8}g_s^{-1}$  & D7 $(\mathbf{1})$
\\\hline
$ \ell_1 + 3 \ell_2 - e_1 - e_2 - e_3$& $R_1 R_2 R_3 \ell_s^{-8}g_s^{-3}$  &$ 7_3^0\, (\mathbf{1})$
\\\hline
$3 \ell_1 + 2 \ell_2 - e_i - 2 e_j - 2 e _k$& $ R_i (R_j  R_k)^2 \ell_s^{-10} g_s^{-2}$ & $5_2^2$B $(\mathbf{3})$
\\\hline
$2 \ell_1 + 3 \ell_2 - e_i - 2 e_j - 2 e _k$ & $  R_i ( R_j R_k)^2 \ell_s^{-10} g_s^{-3}$ & $5_3^2 \,(\mathbf{3})$
\\\hline
\end{tabular}
\end{center}
\caption{Rational curves for $\mathbb{F}^3 \cong$ dP$_4$ and their $d=7$ Type IIB interpretations. Multiplicities are shown in bold. Note that all of the branes above are codimension-2, i.e. they have five non-compact worldvolume dimensions.}
\label{table4}
\end{table}

One may again obtain the (non-)geometric backgrounds in higher dimensions corresponding to each of these by tuning the exceptional curves to zero. Thus for example we conclude that the exotic brane $5_2^2$B descends from a ten-dimensional non-geometric background corresponding to the cycle $3 \ell_1 + 2 \ell_2$, while the uplift of the exotic brane $5_3^2$ corresponds to the cycle $2 \ell_1 + 3 \ell_2$. Recalling that S-duality is implemented by exchange of $\ell_1$ and $\ell_2$, we also conclude that $[5_2^2\mathrm{B}]\stackrel{\,S}{\leftrightarrow} [5_3^2 ]$,  as is known \cite{Bakhmatov:2017les,Fernandez-Melgarejo:2018yxq}. As before, these examples may be generalized by considering further compactifications, leading to the proposal that any curve $\mathcal{C}_{mn}= m \ell_1 + n \ell_2$ corresponds to a class of non-geometric backgrounds, 
\begin{eqnarray}
\label{IIBResult}
\mathcal{C}_{mn}\in H_2(\mathbb{F}^0, \mathbb{Z})  \,\, \leftrightarrow \,\, \left[\left(b_{\mathcal{C}_{mn}}\right)_{n}^{g(\mathcal{C}_{mn})}\right]\,\,\,\,\,
\end{eqnarray}
 labelled by 1-exotic branes as long as it satisfies (\ref{restriction}). We list all possibilities in Table III, indexed by the power  $\alpha$ of $g_s$ in the tension formula, $T \sim g_s^{-\alpha}$. In the current case, we see that $\alpha = n$. 

\subsection{Remarks}
We now remark on the results in Table III. At orders $\alpha = 0,1$ we obviously reproduce the familiar geometric backgrounds corresponding to the fundamental string and D$p$-branes. At order $\alpha=2$, we obtain the NS5 backgrounds $[5_2^0\mathrm{A/B}]$, the KK-monopoles $[5_2^1\mathrm{A/B}]$, as well as three other non-geometric backgrounds. Their existence has already been argued for - they correspond to the 1-exotic branes completing the T-duality orbit $5_2^0$A/B$\stackrel{\,T}{\leftrightarrow}5_2^1$B/A$\stackrel{\,T}{\leftrightarrow} 5_2^2$A/B $\stackrel{\,T}{\leftrightarrow} 5_2^3$B/A$\stackrel{\,T}{\leftrightarrow} 5_2^4$A/B \cite{Kimura:2018hph,Otsuki:2019owg}. 

The non-geometric backgrounds found at order $\alpha = 3,4,5$ capture all 1-exotic branes known in the literature \cite{Bakhmatov:2017les,Fernandez-Melgarejo:2018yxq}. The 0-exotic codimension-0 object $[9_4^0\mathrm{B}]$ has the correct tension to be the S-dual of the D9 brane \cite{Hull:1997kt}. In addition, the del Pezzo curve classification predicts a pair of non-geometric backgrounds $[(-1)_3^8] \stackrel{\,S}{\leftrightarrow} [(-1)_5^8]$ of Type IIB. The $[(-1)_3^8]$ fits together with the known results via T-duality $(-1)_3^8 \stackrel{\,T}{\leftrightarrow} 0^7_3$, though this is not true for $[(-1)_5^8]$ at the level of 1-exotic branes. The representative 1-exotic branes of these two non-geometric backgrounds have not appeared in previous catalogues \cite{Bakhmatov:2017les,Fernandez-Melgarejo:2018yxq,Kleinschmidt:2011vu} since those works studied only exotic branes appearing in dimensions $d \geq 3$, whereas these new 1-exotic branes only appear upon compactification to $d=2$. We are further led to conjecture that these are the \textit{only} additional 1-exotic branes appearing at $d=2$.

Likewise, at $\alpha = 7$ the del Pezzo analysis seemingly allows for a non-geometric background $[0_7^9]$ in Type IIA, corresponding to the cycle $6H + E \in H_2(dP_1, \mathbb{Z})$. This does not correspond to any known exotic brane, nor is it connected via T-duality to any Type IIB non-geometric background which may be labeled by a 1-exotic brane. It appears as a 1-exotic brane only in $d=1$.

We close this section by mentioning that, as in the original work on the M-theory/del Pezzo correspondence, we have imposed by hand a restriction (\ref{restriction}) on the curves of the del Pezzo in order to get a sensible physical interpretation. It would be interesting to see whether the curves violating these constraints could also be given some interpretation in string and M-theory.

\section{Brane Polarization}

In this final section we ask what addition of curves corresponds to in string theory. Given classes $\mathcal{C}_{m_1 n_1}$ and $\mathcal{C}_{m_2 n_2}$, we may use (\ref{IIAResult}) or (\ref{IIBResult})  to obtain the Type IIA/B backgrounds corresponding to the sum $\mathcal{C}_{m_1+m_2,n_1+n_2}$. These formulas imply that objects of tension $T_i \sim g_s^{-\alpha_i}$ add to an object of tension $T \sim g_s^{-\sum_i \alpha_i}$. This is the behavior expected of branes undergoing brane polarization  \cite{deBoer:2012ma}. 

To have brane polarization, one often requires a specific relative arrangement and angular momentum for the branes participating in the process - this data is not yet known to be captured by the del Pezzo. However, the del Pezzo correspondence suggests that if a certain brane polarization \textit{is} possible, it \textit{must} be between a set of branes whose corresponding curves are related by addition. We will require that (\ref{restriction}) be satisfied by the sum $\mathcal{C}_{m_1+m_2,n_1+n_2}$. 

We now give some simple examples involving two branes polarizing to a third. We begin with Type IIA, where the curves have the form $\mathcal{C}_{mn} = m H - n E$. By Table I, one finds
\begin{eqnarray}
\nonumber
\mathcal{C}_{0,-1}+ \mathcal{C}_{1,1} \,\sim \,\mathcal{C}_{1,0}  \hspace{0.18 in} \Rightarrow \hspace{0.18 in} [\mathrm{D0}] +[\mathrm{F1}]\, \sim \,[\mathrm{D2}]\,\,
\end{eqnarray}
This is the usual supertube effect \cite{Mateos:2001qs}. A more exotic example is 
\begin{eqnarray}
\nonumber
\mathcal{C}_{2,1}+ \mathcal{C}_{2,1} \sim \mathcal{C}_{4,2}  \hspace{0.18 in} \Rightarrow \hspace{0.18 in} [\mathrm{D4}] + [\mathrm{D4}]\, \sim \,[5_2^2\mathrm{A}]
\end{eqnarray}
Of course we could also have a polarization involving more than one brane on the right-hand side, e.g. 
\begin{eqnarray}
\mathcal{C}_{2,1}+ \mathcal{C}_{2,1} \sim \mathcal{C}_{3,1} +\mathcal{C}_{1,1}  \hspace{0.05 in} \Rightarrow \hspace{0.05 in} [\mathrm{D4}] + [\mathrm{D4}]\, \sim \,[\mathrm{KK5A}] + [\mathrm{F1}]\nonumber
\end{eqnarray}
These two spontaneous polarizations can be obtained by dualizing the supertube effect, and were discussed e.g. in \cite{deBoer:2012ma}.  On the other hand, note that $\mathcal{C}_{0,-1}$ and $\mathcal{C}_{1,0}$ only add to give $\mathcal{C}_{1,-1}$, which has negative virtual genus. Thus we expect no spontaneous polarization of an isolated D0 and D2 to a third brane.

In Type IIB with curves $\mathcal{C}_{mn} = m \ell_1 + n \ell_2$, we give the following two examples, 
\begin{eqnarray}
\label{IIBexamples}
\mathcal{C}_{0,1}+ \mathcal{C}_{2,1} &\sim& \mathcal{C}_{2,2}  \hspace{0.2 in} \Rightarrow \hspace{0.2 in} [\mathrm{D1}] + [\mathrm{D5}]\, \sim \,[\mathrm{KK5B}]
\nonumber\\
\mathcal{C}_{1,1}+ \mathcal{C}_{1,2} &\sim& \mathcal{C}_{2,3}  \hspace{0.2 in} \Rightarrow \hspace{0.2 in} [\mathrm{D3}] + [\mathrm{NS5}]\, \sim \,[5_3^2]\nonumber
\end{eqnarray}
The first of these gives the Lunin-Mathur geometries \cite{Lunin:2001jy,Lunin:2002iz}. More generally, since we have found that the  KK-monopole always corresponds to the anticanonical class, by (\ref{EMdual}) we conclude that electromagnetically dual branes can always polarize to a KK-monopole. The second process above is discussed in \cite{deBoer:2012ma}.

We thank Eric D'Hoker and Ben Michel for discussions.
\bibliography{Exoticbranes}
\end{document}